\documentclass[sigconf, 10pt, anonymous=False]{acmart}
\usepackage[utf8]{inputenc}
\usepackage{xcolor}
\usepackage{graphicx}
\usepackage{fancyhdr}
\setcopyright{none}
\renewcommand\footnotetextcopyrightpermission[1]{} % 

\title{Cellular LTE and Solar Energy Harvesting for Long-Term, Reliable Urban Sensor Networks: Challenges and Opportunities}
\author{Alex Cabral}
\email{acabral@g.harvard.edu}
\affiliation{%
  \institution{Harvard University School of Engineering and Applied Sciences}
  \city{Allston, MA}
  \country{USA}
  }

\author{Vaishnavi Ranganathan}
\affiliation{%
  \institution{Microsoft Research}
  \city{Redmond, WA}
  \country{USA}
  }

\author{Jim Waldo}
\affiliation{%
  \institution{Harvard University School of Engineering and Applied Sciences}
  \city{Allston, MA}
  \country{USA}
  }

\begin{abstract}
In a world driven by data, cities are increasingly interested in deploying networks of smart city devices for urban and environmental monitoring. To be successful, these networks must be reliable, scalable, real-time, low-cost, and easy to install and maintain---criteria that are all significantly affected by the design choices around connectivity and power. LTE networks and solar energy can seemingly both satisfy the necessary criteria and are often used in real-world sensor network deployments. However, there have not been extensive real-world studies to examine how well such networks perform and the challenges they encounter in urban settings over long periods. In this work, we analyze the performance of a stationary 118-node LTE-connected, solar-powered sensor network over one year in Chicago. Results show the promise of LTE networks and solar panels for city-wide IoT deployments, but also reveal areas for improvement. Notably, we find 11 sites with inadequate RSS to support sensing nodes and over 33,000 hours of data loss due to solar energy availability issues between October and March. Furthermore, we discover that the neighborhoods most affected by connectivity and charging issues are socioeconomically disadvantaged areas with a majority Black and Latine residents. This work presents observations from a networking and powering perspective of the urban sensor network to help drive reliable, scalable future smart city deployments. The work also analyzes the impact of land use, adaptive energy harvesting management strategies, and shortcomings of open data, to support the need for increased real-world deployments that ensure the design of equitable smart city networks.
    
\end{abstract}


\acmYear{2023}\copyrightyear{2023}
\acmConference[SenSys '23]{ACM Conference on Embedded Networked Sensor Systems}{November 13--15, 2023}{Istanbul, Turkiye}

\begin{document}

\maketitle
\pagestyle{empty}
\section{Introduction}

% reliable, infrastructure-less
% can't just put sensors up
% won't just run forever on their own
% how to sustain it

% land use, how to sustain power and connectivity, what to keep in mind for a plug and play network

As the global urban population continues to grow, cities are increasingly interested in monitoring urban processes such as vehicular traffic, and public health and environmental harms including air pollution and noise, to help cities grow in a healthy and sustainable fashion~\cite{desa201868,townsend2013smart,Okafor_Aigbavboa_Thwala_2022}. The lowering cost of sensing infrastructure and recent digital twin capabilities have encouraged city officials, researchers, and urban residents to use large-scale, low-cost sensor networks to monitor hyperlocal phenomena, inform policy and planning decisions, and collect data to help transition to being considered smart cities~\cite{daepp2022eclipse,Okafor_Aigbavboa_Thwala_2022,o2023pursuit,rashid2016applications}. 

We identify that, to be successful, a smart city network must be:
\begin{itemize}
    \item \textbf{reliable}: the network should continue to operate and transmit data over long periods of time and across the city to ensure equitable node distribution~\cite{Kafi2013,murty2008citysense}
    \item \textbf{scalable}: it should be easy to add/replace nodes within the network at any new location in the city~\cite{murty2008citysense}
    \item \textbf{easy to maintain}: nodes should be outfitted with hardware and firmware that minimize the need for in-person maintenance~\cite{dehwah2015lessons,daepp2022eclipse}
    \item \textbf{real-time}: data must be transmitted as quickly as possible, particularly for applications such as emergency services~\cite{Kafi2013}, and the network must be monitored in real-time for maintenance~\cite{murty2008citysense}
    \item \textbf{low-cost}: by using existing infrastructure and services, the network can avoid added costs in installation and maintenance~\cite{murty2008citysense,Kafi2013,daepp2022eclipse}
\end{itemize}

We determine that two key features of an urban sensor network's design can help to make the network fit within the aforementioned criteria. The first is \textit{connectivity}, which is essential for data transmission, real-time node monitoring, and software updates. The second is \textit{power}, which provides for reliable operation and data collection. The decisions that cities and network designers make in these two areas have a direct and significant impact on the criteria for a successful smart city network. For example, an urban sensor network that uses a low-power wide-area network (LPWAN) for connectivity may not satisfy the criteria of low cost because the backhaul infrastructure required, although low in per-unit cost, quickly becomes expensive when considering the number of cells required for a large, dense sensor network~\cite{hossain2021comparison}. Similarly, a smart city network that relies on wired power may not be scalable, as nodes will be limited to locations that already have wired mains~\cite{adkins2018signpost} and will involve additional installation and maintenance cost.

\begin{figure*}
    \centering
    \includegraphics[width=\textwidth]{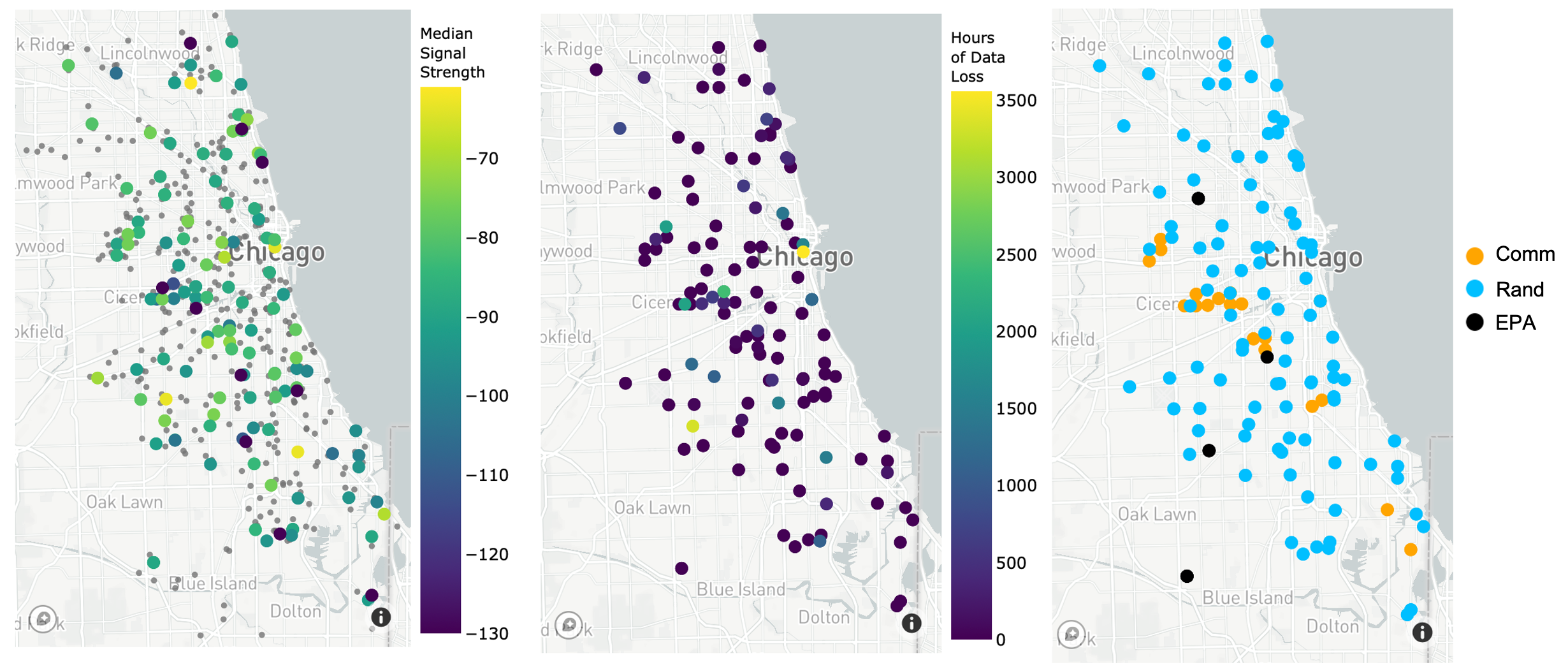}
    \caption{The first map shows the location of each sensor node (large dots) color-coded by its median signal strength and the location of all cellular towers (small gray dots) connected to during the study. Sensor nodes with a median signal strength of -130 (colored in dark purple) are those that could not connect and had to be relocated. The second map shows each sensor node color-coded by the number of hours of data it lost from charging issues in the winter months.
%One of the nodes with the highest number of hours in power saving mode is in downtown Chicago close to several tall buildings, but another node with nearly the same number of hours is in the western part of the city close to Cicero. Thus it is clear that simply considering areas based on the number or density of tall buildings is not enough to determine whether or not they will experience solar charging issues in winter months
The third image shows the node locations color coded by how they were selected---randomly, by community groups, or at EPA stations for calibration}
    \label{fig:network_maps}
\end{figure*}

Based on a review of prior urban sensor network deployments and our experience working on a large-scale sensor network, we establish that LTE networks and solar panels are the appropriate connectivity and power choice for most urban sensor networks given the available options and necessary criteria. Although LTE performance for mobile communication in urban areas is well-researched~\cite{choudhary2020use,lee2018cell,di2019radio}, the performance of IoT-specific networks when implemented in a city-scale long-term sensor network deployment is yet to be characterized. Solar power in urban sensor networks has also been evaluated on a small scale~\cite{dehwah2015lessons}, but not in a large-scale long-term deployment. Moreover, there are no established guidelines that can ensure reliable performance for future deployments of such large-scale LTE-connected, solar-powered sensor networks. Finally, researchers have not looked into the overlap between technical issues that arise in LTE connectivity or solar power and the socioeconomic factors that make up many ``sensor deserts"~\cite{robinson2021sensor}, or areas that lack nodes in cities with sensor networks.

In this work we describe the design and analyze the connectivity and power performance of a stationary 118-node LTE-M connected, solar-powered sensor network deployed for one year in Chicago, Illinois. We find that 11 of the 118 original node locations could not support LTE connectivity, despite all FCC and network provider connectivity maps indicating otherwise. A small number of cell towers and node locations are disproportionately affected by significantly delayed readings, and 44 of the 118 nodes experienced issues charging in the winter months. Furthermore, we discover that connectivity and power related issues are not equitably spread around the city, but rather are more prominent in areas that are classified as socioeconomically disadvantaged and have a larger racial minority population. 

Our primary contribution is an in-depth analysis of a long-term real-world deployment assessing the feasibility and reliability of a large-scale LTE-connected and solar-powered urban sensor network. Additional contributions include: 1) highlighting the overlap between technical challenges in urban sensor networks and socioeconomic inequality, 2) revealing the inherent challenges in relying upon open data sources that are commonly used to predict connectivity and power availability for urban sensor network deployments, and 3) identifying strengths and weaknesses to define future research directions in energy harvesting systems and equitable network infrastructure deployments to ensure the just future of smart city networks.

This paper is structured as follows: Section 2 offers an overview of Related Works; Section 3 highlights why the city of Chicago is a useful case study for urban sensor networks; Section 4 highlights the design of the sensor network and datasets used; Section 5 discusses the connectivity of the sensor network, including the hardware, network carrier information, and insights from the year-long deployment; Section 6 details the powering of the sensor network, including the hardware, energy management techniques, and insights from the deployment; Section 7 provides a discussion, focusing on the implications of the challenges we discovered and the limitations of our study.%; Section 8 is the conclusion.

\section{Related Works}
In this section, we first review former and existing sensor network deployments to identify necessary criteria, prior evaluations, and known issues around inequality. We then examine LTE connectivity and solar power in urban areas, as these are the technologies we use for our sensor network.
\subsection{Criteria for Urban Sensor Networks}
By examining prior urban sensor network deployments, we have identified five criteria necessary for success---reliability, scalability, ease of maintenance, real-time communication, and low cost. The shortcomings of prior sensor networks has often been caused by a lack of reliability, either in terms of not functioning over time, as with malfunctioning hardware~\cite{catlett2017array,daepp2022eclipse,Fang_Bate_2017}, or not communicating data reliably over space and time~\cite{Kafi2013}. Many prior networks have also raised the issue of scalability, which is especially prevalent when relying on electrical cables and wired power, which may be available at street lamps or traffic signals, but ultimately limits the node placement locations~\cite{adkins2018signpost,daepp2022eclipse}. Similar initiatives have shown that reliance on these specific locations can additionally make installation and maintenance more difficult, which then increases the cost of operation~\cite{nyccasStudy,Kafi2013}. The issue of maintenance is particularly important in urban settings, where the cost of accessing a node can be very high~\cite{dehwah2015lessons}.  

Conversely, we find that some deployments are more successful because they achieve low-cost via the use of existing infrastructure. For example, officials in New York City chose to use an existing public safety wireless network for a new traffic control system~\cite{townsend2013smart} and Chicago's Array of Things relied on cellular networks~\cite{catlett2017array}, decisions that helped ease installation and thus save costs. 
\subsection{Evaluations of Urban Sensor Network Deployments}
%Many former evaluations of real-world sensor network deployments focus on indoor environments~\cite{chipara2010reliable} or sensor networks in remote areas, rather than urban ones~\cite{barrenetxea2008hitchhiker}.
The evaluations of real-world sensor network deployments in urban settings have often been small-scale and short-term. A small number of researchers have shared the lessons and challenges learned from urban sensor network deployments, but many of these are focused on specific data such as noise~\cite{offenhuber2020angeles} and water quality~\cite{stoianov2008sensor}. Furthermore, many of these studies rely on the power grid  for high computation tasks~\cite{stoianov2008sensor,catlett2017array}, or use technologies such as Wi-Fi or Zigbee for data transfer~\cite{lee2015low,luomala2015effects,murty2008citysense,rashid2016applications}. The works that evaluate LTE-connected or solar-powered urban sensor networks are small scale and short duration studies that do not offer extended insights on reliability~\cite{luo2020wireless,powar,dehwah2015lessons}.

\begin{figure*}
    \centering
    \includegraphics[width=\textwidth]{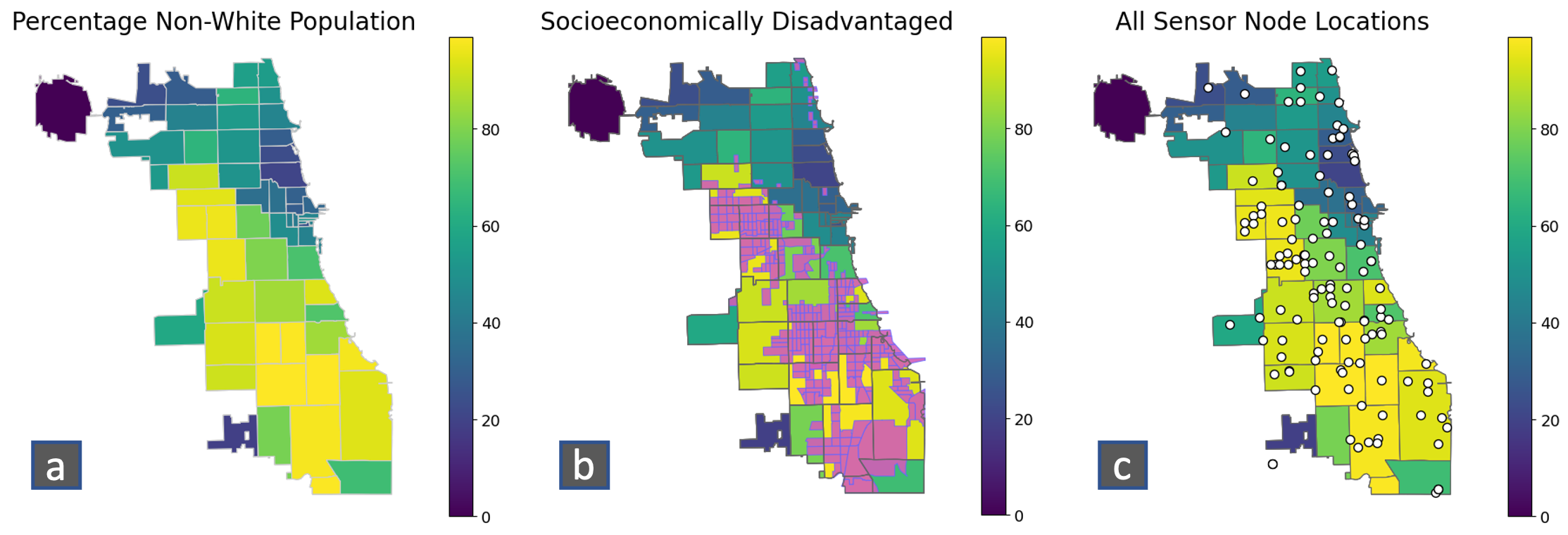}
    \caption{These maps all show the city of Chicago with an outline of each zip code area.%\footnotemark 
    Map a is color coded by the percentage of non-White residents in each zip code, which consists of primarly Black and Latine residents. The maps show that these racial groups live quite separately from White residents, highlighting how racially segregated Chicago is. Map b shows map a with an overlay of the census tracts deemed socioeconomically disadvantaged filled in pink. These areas are selected based on household income, poverty rate, and unemployment rate~\cite{Chicago}. Many of these areas overlap with the zip codes that are majority Black or Latine, highlighting the numerous forms of inequality and segregation present in Chicago. Map c shows all sensor nodes overlaid on map a.}
    \label{fig:socio_maps}
%\end{center}

\end{figure*}

\subsection{Inequality of Sensor Networks}
As smart city networks are increasingly explored and deployed, sociology and urban planning researchers have begun to evaluate the potential social implications of urban sensor networks. For example, one group of researchers evaluated prior urban sensor network deployments and identified areas deemed ``sensor deserts", which are those that lack nearby sensors based on a straight line distance~\cite{robinson2021sensor}. As the researchers state, sensor deserts not only add to existing forms of inequality, but the placement of sensor nodes can also affect resident perception of the distribution of resources and harms throughout a city~\cite{robinson2021sensor}, creating potential political or social strife if nodes are not visible in certain areas. Similarly, others have noted the potential for smart city technologies to ``further deepen the splintering of urban networks, creating deep divides between those with access to 'smart' and those without" and raising questions about the ``politics of urban exclusion"~\cite{townsend2013smart}. Thus, there is an increasing push for equity as a consideration in practical sensor network deployment~\cite{Okafor_Aigbavboa_Thwala_2022,sun2019optimal,Mouton_Ducey_Green_Hardcastle_Hoffman_Leslie_Rock_2019}. 

\subsection{LTE Connectivity in Urban Areas}
Extensive research around mobile connectivity has revealed a variety of factors known to affect RSS and limit propagation distance for LTE signals. These include physical features such as high-rise buildings~\cite{choudhary2020use,lee2018cell}, the distance between the cell tower and receiver~\cite{choudhary2020use,reza2017effect}; meteorological conditions such as precipitation~\cite{choudhary2020use,di2019radio,madariaga2018m}, humidity~\cite{choudhary2020use,luomala2015effects}, strong winds~\cite{choudhary2020use,di2019radio}, temperature~\cite{choudhary2020use,luomala2015effects,sabu2017study} and sudden weather changes~\cite{choudhary2020use,di2019radio}; and environmental measures such as high particulate matter concentrations~\cite{choudhary2020use,lo2019air}. Another major factor that affects signal strength is inter-cell interference (ICI)~\cite{kwan2010survey}, which occurs when a node moves to the edge of one cell tower's range while moving closer to another cell tower. We include all these factors in our analysis of connectivity issues in section 5.

\subsection{Solar Charging in Urban Areas}
Due to the vast quantity of previously deployed solar powered sensor networks and the numerous papers published about these networks, it seems guaranteed that solar power is reliable for most sensor network deployments. However, there have been very few studies looking into the long-term reliability of solar power in urban settings. Dehwah et al.~\cite{dehwah2015lessons} evaluate the performance of a traffic monitoring sensor network in a desert city, and describe the effect of dust storms and building shadows on solar charging. However, they do not do a deep analysis into the locations that were most affected by shadows to determine how the issue may be prevented in future deployments and the potential social implications. 

To our knowledge, this work presents the first in-depth analysis of a large-scale, long-term cellular, solar-powered urban sensor network towards understanding the broader impact of the technical challenges for urban communities. 

\section{Chicago as a Case Study} \label{chicago}
% \footnotetext{The zip code in the northwest (top left) of the city non-residential and encompasses O'Hare Airport, thus does not have demographic statistics. However, it was included in the images for completeness.}.
\subsection{Building Height}
According to the Council on Tall Buildings and Urban Habitat~\cite{council_2023}, amongst cities around the world, Chicago has the 10th most buildings 150 meters and higher, 11th most buildings 200 meters and higher, and 5th most buildings 300 meters and higher. However, its place on those lists is expected to fall within the coming years---Chicago has only three buildings 150 meters and higher under construction and twelve proposed for construction. By comparison, Wuhan, Shenyang, and Bangkok---cities just below Chicago on the list of most 150+ meter buildings---have 49, 14, and 17, buildings under construction respectively, and dozens more proposed in both Wuhan and Shenyang. In addition, development in cities such as Mumbai, Nanning, and Nanjing, which all have several 150+ meter buildings under and proposed for construction will propel them past Chicago in the list in the coming decades. This puts Chicago currently in a unique position for evaluating the impact of built environment towards planning global urban sensor networks.

\subsection{Latitude and Sunlight Hours}
Chicago has a latitude of 41.88 degrees, where the sun is visible for 15 hours, 15 minutes during the summer solstice and 9 hours, 6 minutes during the winter solstice. According to data from the World Economic Forum~\cite{Rae}, the top five most populous latitudes are between the 22nd and 27th parallel north, which are all much closer to the equator and thus have more sunlight on the winter solstice, with an average of 10 hours 35 minutes.

Nevertheless, a number of highly populated cities reside at or above the 42nd parallel north, including London, Moscow, Harbin, and Toronto, as well as much of Western Europe. Cities such as New York and Beijing are also located at nearly the same latitude, receiving  9 hours 13 minutes sunlight on the winter solstice. Furthermore, as the effects of climate change disproportionately affect populations who live closer to the equator, mass migration away from the equator is expected~\cite{Lustgarten_2020}. Thus, understanding the performance of solar-powered sensor networks at northern latitudes is essential for future urban environmental sensing.

\subsection{Segregation and Inequality}
Based on 2020 United States Census Data, Chicago is the fourth most racially segregated large city (population at least 200,000) in the United States~\cite{othering}. Fig.~\ref{fig:socio_maps}a highlights Chicago's racial segregation, showing where the white and non-white---primarily Black and Latine---populations live relative to each other. There is limited data comparing racial segregation in global cities, likely because many countries are more racially homogeneous than the United States. 

However, segregation based on income or social status exists in many global cities, with the highest levels of inequality and segregation often found in cities of lower income countries~\cite{van2021rising}. According to Gini Index data from the 2019 American Community Survey~\cite{Bach_2020}, Chicago has the 10th greatest income inequality amongst US cities, with a Gini index of 0.53 (where a 0 indicates perfect equality and 1 indicates perfect inequality). Compared to cities such as London and Johannesburg, which have the highest global Gini index values---both over 0.7---Chicago has a relatively medium-high level of income inequality~\cite{Chelangat_2019}. As seen in Fig.~\ref{fig:socio_maps}b, the areas of Chicago that are considered most socioeconomically disadvantaged based on factors such as unemployment and poverty level also overlap with many of the areas that have a majority Black or Latine population. Thus, we believe that Chicago provides a useful case study by which to examine the potential social and equity implications that sensing technologies can introduce in cities around the globe.

% \begin{figure*}
%     \centering
%     \includegraphics[width=\textwidth]{figures/chi_socioeco.png}
%     \caption{The first three maps show the city of Chicago with an outline of each zip code area. The zip codes are color coded by the percentage of each major race identified---White, Black, and Latine. These maps show that each of the races is concentrated in different areas of the city, highlighting how segregated Chicago is. The fourth map shows the city of Chicago broken down by census tracts, with the tracts deemed socioeconomically disadvantaged filled in blue. These areas are selected based on household income, poverty rate, and unemployment rate~\cite{Chicago}. Many of these areas overlap with the zip codes that are majority Black or Latine, as shown in the other maps, highlighting the numerous forms of inequality and segregation present in Chicago.}
%     \label{fig:segregation}
% \end{figure*}

\section{Sensor Network and Data}

\subsection{Sensor Network Design}
The sensor network, described in further detail in [blinded]%~\cite{daepp2022eclipse,daepp2023three}
and shown in Fig.~\ref{fig:network_maps}, was designed and deployed to collect air pollution data across Chicago. The network comprised of 118 unique sensor node locations, with 20 nodes allocated to local environmental justice groups for placement according to their priorities, 12 nodes at four EPA stations (3 nodes at each station) for collocation to perform calibration, and the rest placed based on locations chosen through stratified random sampling, as described in NYCCAS~\cite{nyccasStudy}, with a small subset chosen by partner organizations.
% \begin{itemize}
%     \item 80 devices placed based on locations chosen through stratified random sampling, as described in NYCCAS~\cite{nyccasStudy}
%     \item 20 devices allocated to local environmental justice groups for placement according to their priorities
%     \item 12 devices at four EPA stations, 3 devices at each station, for collocation to perform calibration
%     \item 6 devices for partner organizations to address their priorities 
% \end{itemize}

All devices that were not at EPA stations were installed at bus shelters throughout the city, as shown in Fig~\ref{fig:shadow_tree}. These nodes were placed at the same height, about 2.5 meters above ground. Nodes at EPA stations were located on the rooftops near the EPA monitors, several meters above ground and at different heights based on the height of the building or structure housing the EPA monitor. Most of the devices were installed at their respective locations in July and August 2021, with 98 nodes (over 83\%) placed by July 3rd, 2021. %Notably, one set of EPA sensors were installed in April 2022, but their data are still included in the analysis for completeness.

\subsection{Datasets}
The node-related data for each reading, including the time, received signal strength (RSS), battery level, internal node temperature, and air pollutant readings were all logged with each reading and stored in an cloud server. We calculated the latency by comparing the time of the sensor reading to the time of the data's insertion into the server. Cell tower information, such as the cell tower ID, were collected when making a connection with the tower. We used OpenCellID~\cite{opencellid} to link the cell tower information with locations, OSM (Open Street Maps) Buildings~\cite{OSMBuildings} to gather data about buildings surrounding the nodes, FCC Broadband~\cite{fcc} and nPerf~\cite{nPerf_data} data to examine AT\&T connectivity, Meteostat~\cite{meteostat} to collect external weather data, and the Shadow Accrual Maps tool~\cite{nyushadow} to calculate the amount of shadow hours at each node location. Socioeconomic data were pulled from the City of Chicago Open Data Portal~\cite{Chicago}. 

\subsection{Data Cleaning}

We removed readings that had no connectivity data (N = 9,393, 0.2\% of readings), readings where the signal was equal to zero (N = 11,626, 0.12\%), readings where the tower location was clearly outside of Chicago, possibly due to sensors being shipped back and forth when there were issues (N = 11,778, 0.12\%), and readings with a delay of more than 24 hours (N = 54,900, 0.63\%), as this was likely indicative of a device issue, rather than connectivity or charging issue. We also identified 565,371 readings (12.7\%) where the cell tower could not be located in the OpenCellID database; we kept these readings in for all analyses except ones involving distance and general direction of the cell tower.

\section{Connectivity}

\subsection{Motivation for an LTE-Connected Urban Sensor Network}

Despite recent advances in WiFi and low-power wide-area networks (LPWAN), such as LoRaWAN~\cite{haxhibeqiri2018survey,khalifeh}, most urban sensor networks will rely on cellular networks in the coming years %due to widespread global availability and the lower cost and ease of setup and scaling.
for the following reasons: 1) Dependence on existing urban cellular networks ensures city-wide coverage without additional infrastructure. 2) Widespread global availability and flexible data plans with each generation. 3) Lower cost and ease of setup and scaling---for technologies such as LoRaWAN, scalability is a particularly pressing issue due to the cross-technology interferences that will arise from other technologies~\cite{haxhibeqiri2018survey} and potential packet collisions with large sensor networks~\cite{hossain2021comparison,lavric2018performance}. In addition, LPWAN require dedicated infrastructure that have a low per-unit cost, but quickly add up in costs based on the cells required to support high node density~\cite{hossain2021comparison}. 

Thus, to support the necessary criteria of reliability, real-time, and low cost, we use an LTE network for communication. LTE networks propose great coverage in most cities around the globe~\cite{fcc, nPerf_2023}, providing means for scaling reliably. Because the cellular infrastructure is already built and evolving, networks are easy to set up and remain low-cost, especially with the variety of LTE plans available. Finally, with the fast evolving generations of cellular communication, such networks are increasingly seen as dedicated low latency connectivity for massive IoT deployments in growing cities~\cite{qadir2023towards}. 

\subsection{Materials: Antenna and LTE Carrier}
\begin{figure}
    \centering
    \includegraphics[width=\linewidth]{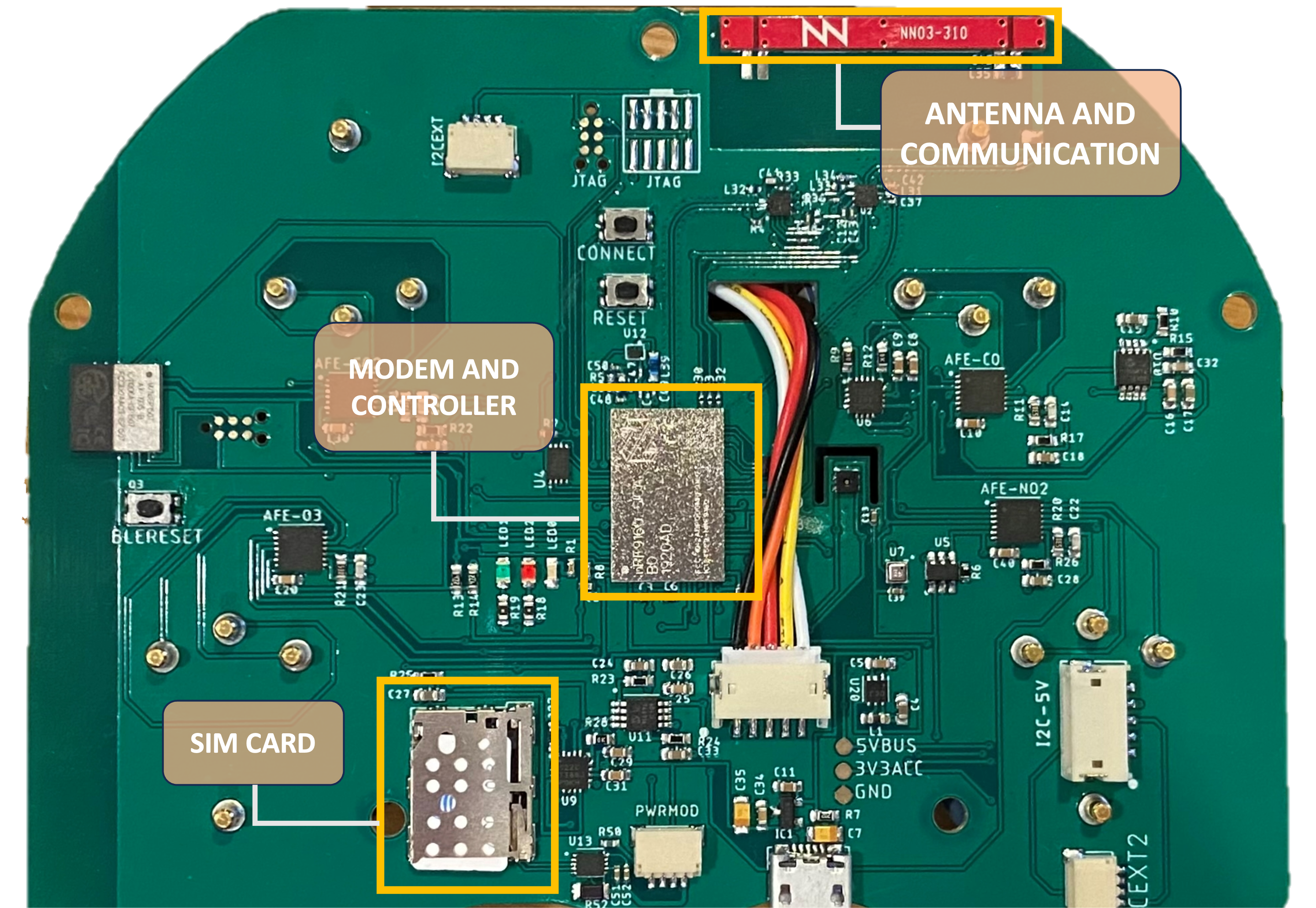}
    \caption{The sensing nodes's printed circuit board (PCB) with connectivity hardware highlighted.}
    \label{fig:pcb}
\end{figure}

The sensing nodes connected via AT\&T's 4G IoT LTE-M One network, which uses LTE Bands 2, 4, and 12, and operates at frequencies of 700, 1700, and 1900 MHz. Each node used a SIM card and Ignion NN03-310 antenna~\cite{Ignion_2022}, which transmits data over 3G and 4G, is tuned for channels 2, 3, 4, 5, 9, 12, 20, and 28, and operates on frequencies from 698-960 MHz and 1710-2690 MHz. The antenna was placed at the top right of the printed circuit board (PCB)~\footnote{After conversations with the antenna manufacturer and a small series of tests, it was determined that antenna placement on a PCB can have a significant effect on the RSS values. It is imperative for sensing node designers to consult with antenna manufacturers to ensure correct antenna placement on custom PCB for the best connectivity.}, as shown in Fig~\ref{fig:pcb}.

\subsection{Methods: Node Connectivity and Data Transmission} \label{data}

The sensing node preserved battery life by periodically waking up to record a sample and transmit data to the cloud, as further described in Section~\ref{power}. For this deployment, the nodes were set to transmit data every five minutes from the last recorded sample time. The data transmission process included the following series of steps: 1) The microprocessor woke up and kicked off two processes on separate threads, 2a) One thread sampled the sensor with the longest latency, typically about 8 seconds, 2b) A separate thread simultaneously initiated connection to the cloud, 3) Another array of low latency sensors were sampled, 4) The data were then packaged and transmitted to the IoT endpoint going through the cell tower, AT\&T network routers etc.
% \begin{enumerate}
%     \item The microprocessor woke up and kicked off two processes on separate threads
%     \item The two threads run:
%     \begin{enumerate}
%         \item One thread sampled the sensor with the longest latency, typically about 8 seconds
%     \item A separate thread simultaneously initiated connection to the cloud
%     \end{enumerate}
%     \item Another array of low latency sensors were sampled
%     \item The data were then packaged and transmitted to the IoT endpoint going through the cell tower, AT\&T network routers etc.
% \end{enumerate}

\subsection{Methods: Retry Logic}
If a node could not connect to the cloud, it stored the reading locally, went back to sleep for five minutes, and tried to connect again. After 10 retries, if the node still could not connect, then the node was set to reboot itself. After a reboot, the node would immediately try to make a connection to the cloud and would not record local readings until it did because the node lacked a real time clock. Once the node could connect again, it transmitted all locally stored data and errors that were logged in the absence of connectivity.

\subsection{Results: Readings and Cell Towers}

For the one-year period and 118 nodes in our network, our dataset included 8,684,756 readings. We linked the readings to 417 unique cell tower locations, 65 with only 1 associated reading, 179 with 500 (0.0057\%) or more readings, and 165 with 1000 (0.011\%) or more readings. %median is 153 readings (0.00175\% of all readings). 

\begin{figure*}
    \centering
    \includegraphics{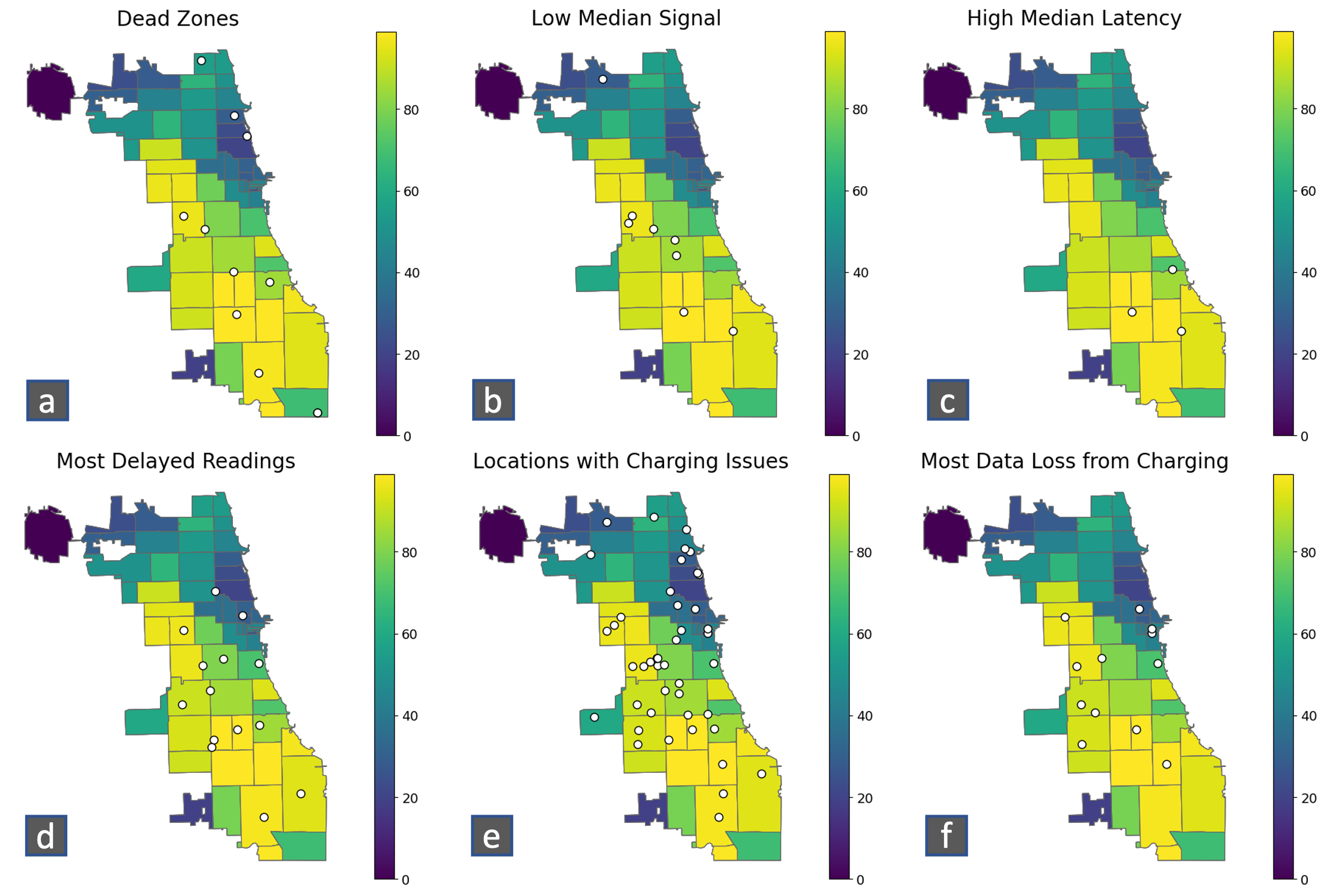}
    \caption{These maps show the locations of sensors with connectivity or charging issues, all overlaid on the city of Chicago with each zip code colored based on its percentage of non-White residents. Map a shows the ``dead zone" locations where the nodes could not connect to the LTE network. Map b shows the locations where is sensing nodes had a median signal $\leq$ 100 dBm. Map c shows the locations where sensing nodes had a median latency $\geq$ 5 seconds. Map d shows the locations with the most delayed readings. Map e shows all of the node locations that experienced solar charging issues in the winter months, and map f shows the node locations with the most data lost from solar charging issues, all with over 1100 hours of data loss.}
    \label{fig:sensor_maps}
\end{figure*}

\subsection{Results: ``Dead Zones"}

Over the course of our deployment, we identified 11 locations (9.32\%) at which the sensor nodes reported consistently low RSS values and ultimately failed to connect, generally within a few days of installation. These 11 locations include 10 from the main deployment beginning in July 2021 and one node location from an earlier pilot program in April 2021. 3 of the 11 locations were selected for deployment by local community groups, a significant percentage more than in the overall deployment. Initial mitigation strategies involved moving the nodes to the closest bus shelter, which was often directly across the street. However, we discovered that the nodes had to be moved even further---sometimes multiple blocks away---to establish a connection. 

\begin{figure}
    \centering
    \includegraphics[width=\linewidth]{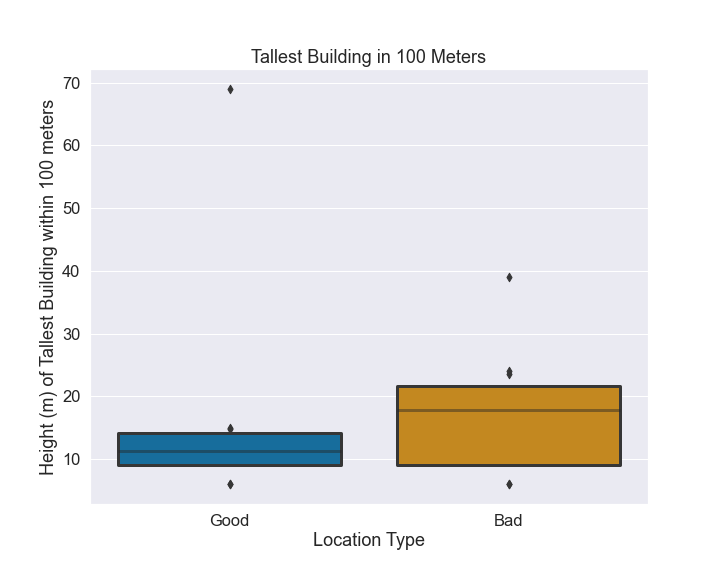}
    \caption{This boxplot shows the height of the tallest building within 100 meters of a node location at its "Bad" location at a dead zone, and its "Good" location at the new bus shelter after the node was moved. The distribution of building heights is generally taller at the dead zone locations, suggesting a correlation between nearby building height and connectivity.}
    \label{fig:badvsgood}
\end{figure}

We examined a number of factors to determine the potential cause of these ``dead zones", including the distance between the node and cellular tower, the number of towers close to a node, evidence of inter-cell interference (ICI)~\cite{kwan2010survey}, and nearby physical urban structures, including the distance and height of the closest building to the node, and the number, tallest height, mean and median building height within 100, 250, and 500 meters of each node. We found no evidence to suggest that any of these features had an effect on a node's ability to connect, when comparing all ``dead zones" to all other node locations. When comparing ``dead zone" locations to the new locations each of those nodes was moved to, we found a statistically significant difference in the height of the tallest building within 100 meters of the node after relocation versus before, as shown in Fig.~\ref{fig:badvsgood}. This indicates that land use and urban form close to the location of stationary sensors are likely factors impacting connectivity, fitting in line with observation from prior work~\cite{choudhary2020use,lee2018cell}.

\begin{figure}[tb]
    \centering
    \includegraphics[width=0.43\textwidth]{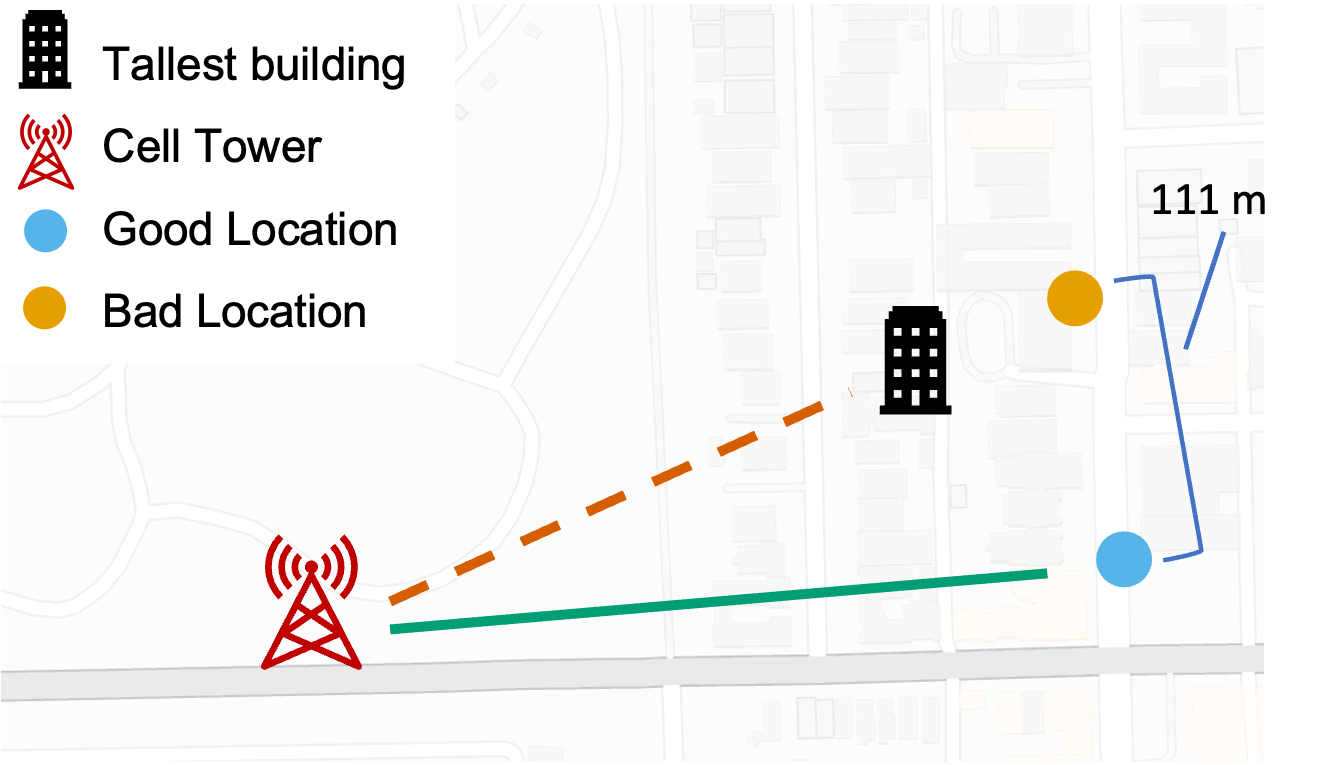}
    \caption{This map highlights the line of sight issue for one of the nodes originally placed in a dead zone then moved to a new location, where it could connect to the same LTE tower it originally had trouble connecting to. The building is the tallest building within 100 meters of the dead zone, with a height of 24 meters. %This building is 24 meters tall and 37 meters away from the dead zone. The tallest building within 100 meters of the new location is 12.4 meters.
    }
    \label{fig:lineofsight}
\end{figure}

In addition, we investigated the role of line-of-sight as a primary factor contributing to ``dead zones". We examined the relation between the sensor node, cellular tower, and tallest nearby building for the two nodes found to connect to the same primary cellular tower at their original (``dead zone") and new location. We found that one of these node configurations exhibited line-of-sight interference, as shown in Fig.~\ref{fig:lineofsight}, as the tallest building (11.9 meters) was clearly in the path between the cellular tower and sensing node. %A second node, which connected successfully only once from its original location, also had a taller (17.8 meters) building within 100 meters in the dead zone than at its new location, though this building did not seem to block the line of sight between the node and tower based on the recorded location. Clearly 
Due to the limited number of examples to examine, there is a need for further investigation in larger datasets, however, this evidence supports the key role of line-of-sight impediments in contributing to ``dead zones". 

Finally, we examine the socioeconomic factors around the node locations without connectivity. We do not find a significant difference in the socioeconomic factors when comparing node locations that can and cannot connect, likely because there are a large number of nodes around the city. However, we do note that many of the dead zone locations are in socioeconomically disadvantaged and majority Black and Latine neighborhoods, as shown in Fig.~\ref{fig:network_maps}a.

\subsection{Results: Signal Strength}

As shown in Fig.~\ref{fig:network_maps}, the yearly median signal strength for each node ranged from -61 dBm to -113 dBm, with a network-wide median of -87 dBm. There was no significant difference in the median signal strength for community-selected versus randomly-selected nodes and we did not identify a statistical relationship between surrounding physical features, such as building height or distance to buildings, and the median signal strength for the sensor node or corresponding cell tower location.

As with ``dead zones", we found that the node locations with the lowest median signal strength---those less than 100 dBm---were nearly all sited in neighborhoods that are socioeconomically disadvantaged and have a higher percentage racial minority population. In fact, only one of the eight locations with a low median signal strength was sited in a majority white neighborhood, as shown in Fig.~\ref{fig:socio_maps}b.

\subsection{Results: Latency}
%To estimate the latency of each reading, we compared the time of the reading on the sensing node to the time the data were inserted into the Azure server. 
We found that over the entire year's worth of data, the minimum latency was 2 seconds, the median latency was 5 seconds, and the interquartile range fell between 4 and 6 seconds (our data allowed only for estimating seconds, and not milliseconds for latency). 

When examining the median latency for each sensor node over the course of the study, we found a much tighter distribution then we saw for median signal strength. In fact, the interquartile range all falls at the exact same value of 5 seconds. There are only three sensor locations with a median latency greater than that value, shown in Fig.~\ref{fig:network_maps}c, and two of those locations overlap with those that have poor median signal strength, suggesting a correlation between signal strength and latency. %Looking at the map of the sensor locations in Fig~\ref{fig:nonwhite_maps}, we see again that most of these locations are in neighborhoods with a majority Black or Latine population. 

\begin{figure}
    \centering
    \includegraphics[width=\linewidth]{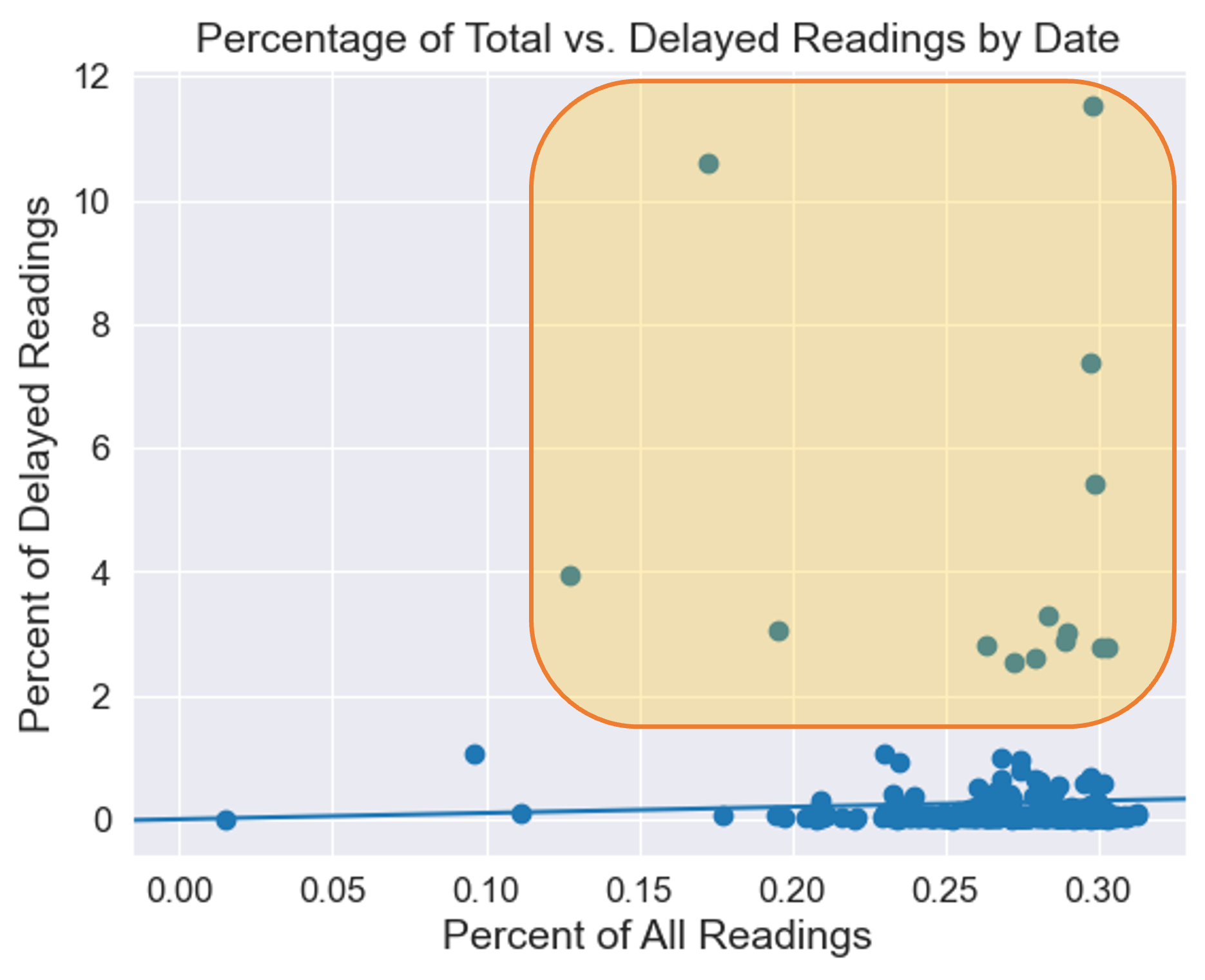}
    \caption{This scatterplot displays the percentage of readings with $\geq$ 30 seconds latency compared to the percentage of all readings by date, showing that significantly delayed readings tend to occur on specific dates.}
    \label{fig:percentages}
\end{figure}

We find that only 7.24\% of readings have a latency of 10 or more seconds, 1.18\% have a latency of 30 or more seconds, and less than 1\% (0.88\%) have a latency of one minute or longer. Although these are low percentages, we examined the significantly delayed readings to determine if they occur randomly or follow a pattern. We found that the delayed readings do not occur randomly, but rather appeared disproportionately on certain dates, at certain sensor locations, and with certain cellular towers, as seen in Fig.~\ref{fig:percentages}. Interestingly, the sensor locations with the most delayed readings have no overlap with the locations that have either the lowest median signal strength or the highest median latency. However, when looking at the map of the sensor locations in Fig~\ref{fig:socio_maps}d, we see again that most of these locations are in neighborhoods with a majority Black or Latine population. We could not identify any temporal or location-based events events, such as sporting games, that have previously been associated with cellular network delays and may have caused these significant events. Coupled with the lack of empirical evidence from the cellular service providers% and confirmation from the cellular carrier
, we are led to determine that the delays are likely caused to carrier-specific issues such as cell tower maintenance.

% \begin{table}[]
%     \centering
% \begin{tabular}{|c|c|c|}\hline 
%     Min. Latency (s) & No. of Readings & Pct. of Readings \\ 
%     \hline\hline
%     7 & 1,509,068 & 17.38 \\
%     10 & 628,416 & 7.24 \\
%     15 & 252,385 & 2.91 \\
%     20 & 155,340 & 1.79 \\
%     30 & 102,047 & 1.18 \\
%     60 & 76,471 & 0.88 \\
%     120 & 68,429 & 0.79 \\
%     300 & 63,135 & 0.73 \\
%     3600 & 34,150 & 0.39 \\
%     7200 & 22,594 & 0.26 \\ \hline
% \end{tabular}
% \caption{This table shows the number and percentage of readings from the data set with a specified minimum latency. The vast majority of readings were transmitted from the node to the Azure server within 30 seconds, showing the promising potential of LTE-connected sensor networks.}
%     \label{tab:latencytable}
% \end{table}

% \begin{figure*}[tb]
%     \centering
%     \includegraphics[width=\textwidth]{figures/threePercentages.png}
%     \caption{These scatterplots show the percentage of readings with at least 30 seconds latency compared to the percentage of all readings by date, tower location, and sensor location. From these plots, it is clear that the significantly delayed readings tend to occur on specific dates, at specific sensor node locations, and with a small number of cellular tower locations.}
%     \label{fig:percentages}
% \end{figure*}

% \begin{figure}
%     \centering
%     \includegraphics[width=\linewidth]{figures/nobuilding.png}
%     \caption{This bus shelter is at a location reported to have shadows based on the NYU shadow tool and open data. However, it is clear that there is no building and are no shadows impeding any kind of solar charging, revealing a challenge in relying on open data.}
%     \label{fig:nobuilding}
% \end{figure}

\section{Power}
\subsection{Motivation for a Solar-Powered Urban Sensor Network}
Nodes must be continuously running to collect data over time, yet many outdoor urban spaces are not equipped with accessible wired mains~\cite{adkins2018signpost}. Solar power is the most ubiquitous form of renewable energy for sensor networks, and will remain prevalent in the coming years for the following reasons: 1) Solar panels are relatively inexpensive and easy to install. 2) Solar panels can power sensors that need to operate continuously in remote or hard-to-reach locations where it may be difficult or expensive to run electrical cables or replace batteries. 3) Using solar power eliminates the need for frequent battery replacements, which creates an added burden for cities looking to deploy sensor networks. %4) Solar power is environmentally friendly and reduces the carbon footprint associated with operating the sensors, which is desirable for cities looking towards a sustainable future.

Thus we use solar energy to power our sensor network to  %a solar-powered sensor network thus helps network designers 
achieve reliability through continuous power, scalability in allowing for power in locations that do not have outlets, ease of maintenance by limiting battery replacements, and low-cost by requiring no new infrastructure. %For these reasons, we use solar power as the energy source in our sensor network. 

\subsection{Materials: Battery, Solar Panel, and Power Usage}~\label{power}
Each sensing node was outfitted with a rechargeable 2000 mAh lithium polymer battery%, which was charged using 
 and a 10$\times$13 cm Voltaic Systems P126 6W solar panel. The solar panel was attached horizontally, in a flat position, to the top of the node's respective bus shelter to maximize solar absorption, maintain security of the panel, and provide ease of installation.

To optimize for low power consumption, the microcontroller operated in a duty cycled mode, consuming as little as 40~{\textmu}A between measurements. The device's four electrochemical gas sensors consume microwatts of power, while the particulate matter (PM) sensor consumes up to 80~mA power as it relies on an internal fan to circulate air. Thus to optimize the overall power usage, we sampled the gases every 60 seconds and sampled the PM and transmitted data every 5 minutes. On average, the device drew 4mA current over a 24 hour period, allowing the battery to power the sensing node, including communications, for approximately 15 days at the aforementioned sampling rate.

\subsection{Methods: Power Saving Strategies}
In October 2021, we noticed that one of the devices was no longer charging. After sending the local maintenance team to investigate, we discovered that the sun was no longer reaching the solar panel due to the change in the sun's position and the node's location surrounded by skyscrapers. We anticipated that this issue would begin to show up in other nodes as well, so determined three potential solutions to ensure the network still collected useful data throughout the winter months:

\begin{enumerate}
    \item Set the sampling interval to be more than every five minutes, which would deplete the battery less quickly by running the PM sensor and data transmission less often.
    \item Implement a power-saving mode to ensure devices only run when they have a certain amount of battery and sleep when they are below that value.
    \item Schedule devices to only run at certain times of the day, i.e. for a few hours in the middle of the day when there is sunlight.
\end{enumerate}

Naturally, each option comes with its own trade-offs that had to be considered. Sampling less often would provide less temporal coverage which could cause cities to potentially miss timely notifications from sensors, make it more difficult to identify noisy or anomalous readings through techniques such as moving averages, and introduce calibration errors from datasets with different resolutions. A power-saving mode could result in large time spans with no data, creating difficulty in comparing data from different seasons and potentially resulting in a lack of data needed for calibration. Scheduling devices to only run at certain times would limit data collection to only specific hours of the day, and may not solve the issue if the number of hours is not chosen correctly. 

Based on the tradeoffs and our need of data for sensor calibration, we implemented a power-saving mode to put devices into a deep sleep to avoid depleting the batteries in low- or no-light conditions. Power-saving mode was initiated when a battery's power level fell to 15\% or less of its total capacity then turned off when the battery's power level had recharged to at least 40\%.

\subsection{Results: Data Loss due to Power Saving Mode}

Between the autumn and spring equinox of the year long study period, 44 devices (37.29\%) went into power saving mode (PSM), with most devices entering PSM between January and March. Seven of these devices were at community selected sites, representing about 16\% of the devices in PSM, indicating the community selected sites were not disproportionately affected. In total, devices in the networks spent 19,450,915 seconds --- over 33,180 hours or 1382.5 days---in PSM, resulting in about 398,000 potential sensor readings that were not captured. Most devices entered PSM numerous times, with several entering more than five times during the study period. Thus, in many locations there was adequate sunlight to keep the devices charged throughout the winter months if a larger solar panel had been used or the devices had better energy harvesting to extend the battery life with the limited charge they received.

% \begin{figure}
%     \centering
%     \includegraphics[width=\linewidth]{figures/powersavingtimes_barplot.png}
%     \caption{This bar plot shows the number of times that each device entered power saving mode. Many devices entered power saving mode numerous times, indicating that they were able to recharge, even throughout the winter months. Thus one potential solution to address this issue is to have larger solar panels or better energy harvesting for sensor nodes to continue reporting data throughout the winter months.
%     }
%     \label{fig:PStimes}
% \end{figure}

\subsection{Results: Location of Solar Charging Issues}
As expected, the node locations in downtown Chicago entered PSM for a long duration of the winter due to the high number of very tall buildings in the neighborhood. However, several node locations in neighborhoods outside of downtown Chicago, that lack a high density of tall buildings, also experienced solar charging issues. In fact, the node location with the second highest amount of time spent in PSM was not in a location near tall buildings, and 8 of the 12 node locations that had the most power saving hours were outside of the downtown area, as shown in Fig.~\ref{fig:socio_maps}f. The figure also shows that they mostly fall in neighborhoods with a majority Black or Latine population. As seen in Fig.~\ref{fig:shadow_tree}, shadows from trees for large portions of the day could be a potential cause for charging issues in some areas. In addition, ice build up on solar panels may cause charging issues, but this is difficult to diagnose without visiting every node location while it is in PSM. Thus, further analysis is required to determine the exact cause of charging issues in these locations that obviously lack tall buildings in the vicinity. The important takeaway is that the dynamic physical environment of solar IoT deployments need to be considered by tools that are currently being developed to estimate solar energy availability using historic data or satellite/map images~\cite{buchli2014towards,yamin2021online}.

\subsection{Results: Predicting Solar Charging Issues}
\begin{figure}
    \centering
\includegraphics[width=\linewidth]{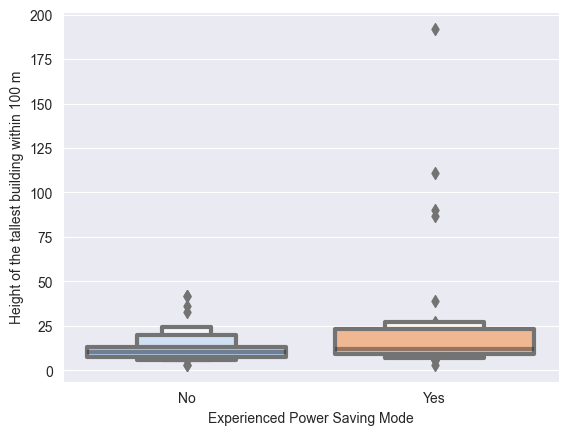}
    \caption{This boxplot shows a comparison of the height of the tallest building within 100 meters for node locations that experienced PSM to those that did not. Excluding the four buildings taller than 75 meters, the distribution of building heights is virtually the same for nodes that did and did not experience PSM, showing that access to building height is not always adequate to predicting solar charging issues.}
    \label{fig:buildingheight}
\end{figure}

We used the OSM Buildings data~\cite{OSMBuildings} and Shadow Accrual Maps tool~\cite{nyushadow} to determine how well we would be able to predict a sensor location having power saving issues. With the OSM Buildings data, we examined the distance to the closest building, height of the closest building, and mean and median height of buildings within 100, 250, and 500 meters of each node location. For shadows, we used the tool to calculate the amount of time each node location was in shadow on the winter equinox. Using both a logistic regression model for the binary case of power saving or not, and a linear regression model for the amount of time spent in PSM, we found no statistical significance for either the amount of time spent in shadow, or any data related to buildings around the node locations, as highlighted for one data point in Fig.~\ref{fig:buildingheight}.

\begin{figure}
    \centering
    \includegraphics[width=\linewidth]{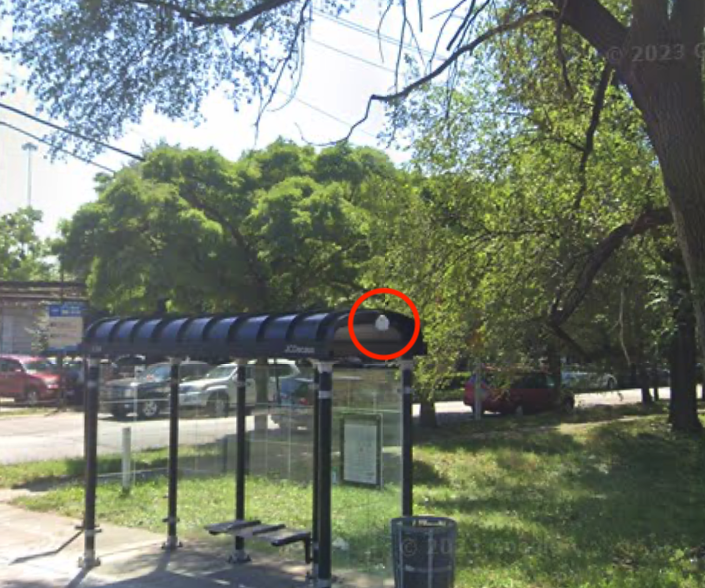}
    \caption{This bus shelter, with the sensor shown in the red circle, is at a location that experienced power saving issues despite having no shadows recorded in the Shadow Accrual Maps~\cite{nyushadow}. It is likely that the large tree blocked the path between the sun and solar panel for several days in the winter. This highlights a challenge in having information about all of the objects that may block solar charging sensors in urban areas.}
    \label{fig:shadow_tree}
\end{figure}

Upon further examination, we discovered that one of the issues around using crowdsourced and open source resources is that they are not consistently updated. For example, one sensor node that was indicated to have shadow issues but did not enter PSM likely had a building present when the data were uploaded, but no longer has a building there as discovered on Google Maps. Likewise, as seen in Fig.~\ref{fig:shadow_tree}, a node location with no building nearby that entered PSM was likely affected by the presence of a tree near the bus shelter, which was not captured in the tools we used, which are focused on buildings. This points to an additional shortcoming of the data available, which focus on buildings and do not account for foliage, hyperlocal snowfall, and other physical phenomena that may impede solar charging.% of solar panels.

\section{Discussion}

\subsection{The Potential of LTE-Connected, Solar-Powered Urban Sensor Networks}

The results show immense promise for LTE-connected urban sensor networks. Most node locations had adequate signal strength to achieve connectivity, and the vast majority of sensor readings were transmitted to the cloud server within five seconds. Furthermore, there were no noticeable issues around connectivity due to temporal features such as weather or traffic patterns. We also had success using LTE to detect errors and perform software updates, including a firmware patch to add the power saving mode. These findings all point to the potential of LTE in creating reliable, scalable, easily maintainable, and real-time sensing in cities. 

Solar panels proved to be a reliable energy source for over half of the year-long study, and most devices that experienced charging issues only did so between January and March. Chicago is at a more northern latitude than most of the global population, so we expect that many cities, and especially those in the Global South, would experience fewer solar charging issues. Additional improvements with solar panel efficiency~\cite{green2014emergence} and research on smart power management strategies for renewable energy in IoT establish solar charging as a viable powering option. 

The nodes that were collocated at EPA stations all experienced no charging or connectivity issues, suggesting that placing nodes on rooftops could be a viable solution to improve reliability. However, node placement is highly dependent on the application, and many cities may choose or need to place nodes closer to street level. Future research could include interpolation and machine learning techniques to correlate data from street level to rooftop nodes to address the technical issues and still collect useful data. Additionally, passive wireless reflector and relay research can find application in routing network availability from cell towers and around built infrastructure to end devices.

\subsection{Implications of Connectivity and Charging Issues}
\begin{figure}
    \centering
    \includegraphics[width=\linewidth]{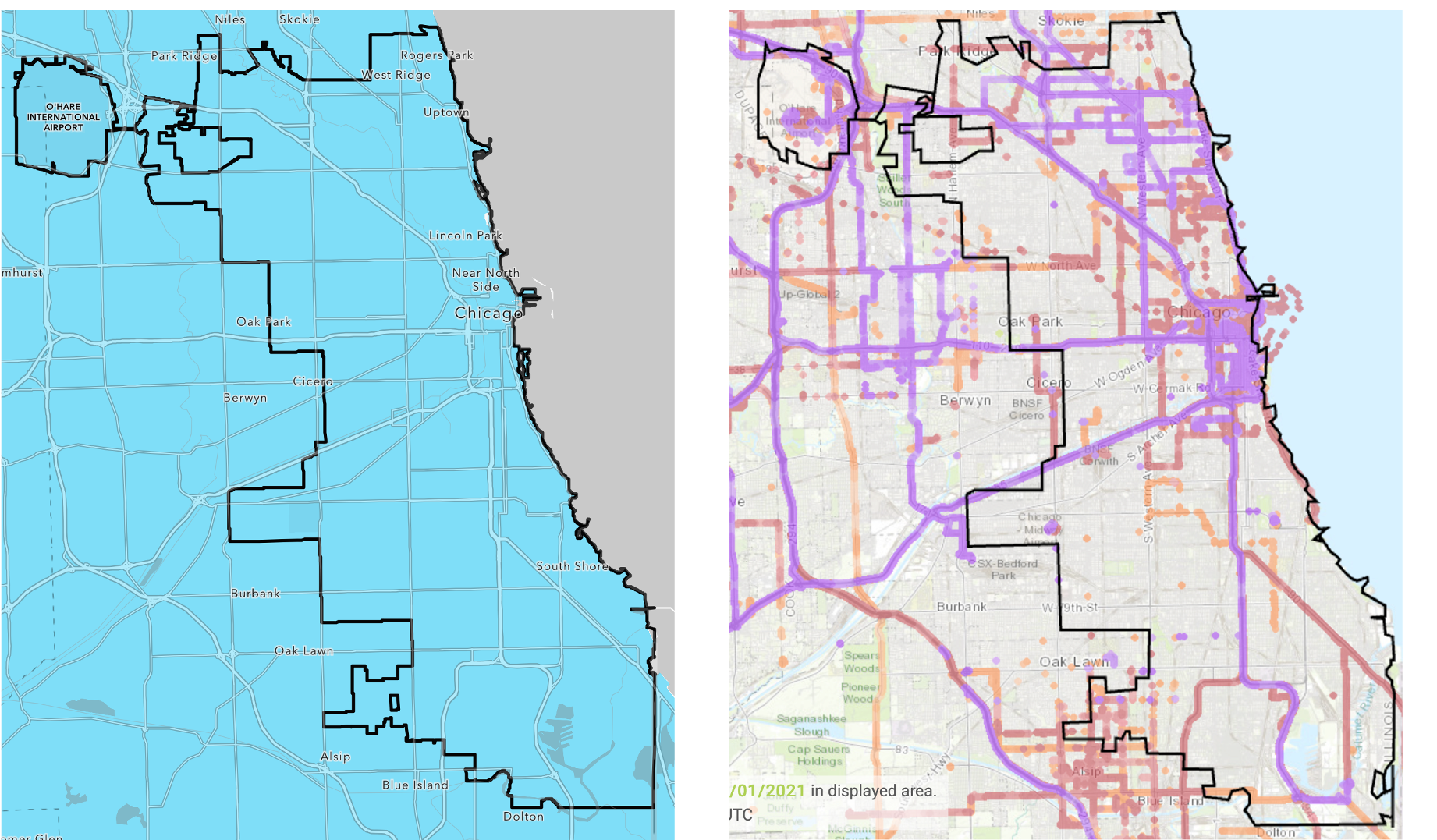}
    \caption{These maps show the FCC (left) and nPerf (right) data for AT\&T LTE in Chicago. The FCC data, in blue, indicates that all of Chicago and its surroundings should have connectivity. The nPerf data, the largest crowdsourced network information, is mostly collected from highways and the northern part of the city, making it incomplete for citywide analyses.}
    \label{fig:carrier_maps}
\end{figure}

Despite the success we had in using 4G LTE-M to transmit data, we discovered issues around ``dead zones", delayed readings, and unequal signal strength. The cause of these issues could not often be easily identified and data sources from AT\&T and the FCC indicate widespread support of the LTE network across Chicago, as seen in Fig.~\ref{fig:carrier_maps}. Thus, the discovery of these issues raises questions on the reliability of LTE networks, especially in cities that do not have as much cellular infrastructure as Chicago. 

However, we did not identify significant data loss from the connection-related issues, suggesting that LTE-connected sensor networks are likely appropriate for applications that do not rely on instant or near instant data. %These applications may be difficult to identify though, because sensor network data may be used for unexpected purposes~\cite{daepp2023three}. 
For applications that cannot afford to have any delayed data, such as emergency support services, network designers will want to think about building robustness into the system to ensure real-time communication for all readings.%, as relying on alternative communication technologies will likely bring about the same or new issues~\cite{}. 

%\subsection{Implications of Solar Power Issues}
Despite the ubiquity of solar panels as the power source for wireless sensor networks, we found that they are not a reliable power source for urban sensor networks for cities %at more northern or southern latitudes 
that have limited sunlight 
%and thus will experience power issues 
in winter months. In addition, urban areas at latitudes closer to the equator will also experience solar charging issues if they have numerous tall buildings blocking the path of the sun. Thus, we need to continue research in alternative charging options, energy harvesting techniques, and battery-less sensors to ensure reliability and scalability in powering urban sensor networks.

%\subsection{The Social Impact of the Challenges}
In our study, we found that cellular connection and solar charging issues are not all localized to areas with tall buildings and may be spread inequitably around a city. Thus, urban sensor network deployments have the potential to exacerbate existing societal inequalities by allowing for networks to be scaled more easily in some neighborhoods than others. In turn, this can increase mistrust between residents and governments~\cite{Okafor_Aigbavboa_Thwala_2022} and drive residents to make assumptions about the distribution of resources and harms based on the physical presence of sensors~\cite{robinson2021sensor}. Thus, to serve people in all communities, sensor network designers should consider working with local service providers, using repeaters, multiple sensors, and other technologies to improve reliability in underserved areas. Furthermore, networking researchers and designers need to focus on equality, and not just quality or area coverage when building and deploying infrastructure. 

\subsection{Challenges around Data Access}
Due to the lack of official up-to-date building information, we relied on open crowdsourced data to determine the location and height of buildings in the city. Similarly, because the location of cellular towers is not publicly available, we relied on data from OpenCellID. As with many open crowdsourced datasets, these data were not completely accurate or up-to-date~\cite{syafiq2016automated}. This was especially clear when examining FCC carrier connectivity information, as the entire city of Chicago seemingly has coverage (Fig.~\ref{fig:carrier_maps}, yet we found that was not the case, likely because the data are reported by carriers~\cite{fcCommission_2023}. We also discovered data accuracy issues in shadow prediction using the Shadow Accrual Maps~\cite{nyushadow}. Other crowdsourced data, such as nPerf, presented an alternative usage issue in incompleteness, as seen in Fig.~\ref{fig:carrier_maps}. Particularly in Chicago, there is significantly more data available in the northern part of the city and along highways, likely attributed to the increased usage of crowdsourced platforms by white people and high-income earners~\cite{chakraborty2017makes,brabham2008moving}. Thus, relying on crowdsourced data makes it difficult to predict locations with solar charging or connectivity issues that may arise due to building height and other urban interferences, made further difficult by the social inequities that exist in many cities and are exacerbated in crowdsourced technologies.%., such as wireless connectivity~\cite{choudhary2020use}. 

The difficulty in working with open crowdsourced data points to a need for new methods to obtain up-to-date %cell tower and building 
urban data. For example, researchers can help develop ways to obtain building height or cell tower location from satellite imagery or Google Maps. We may also look to develop easier ways for cities to create their own databases that are kept up-to-date or develop better community science incentives to keep crowdsourced data sources such as OSM Buildings, OpenCellID, and nPerf up-to-date and to reach new users who do not currently contribute to these datasets. 

\subsection{Limitations of this Study}
We acknowledge that this work is limited, as it focuses on a single-city case study. Although we believe that Chicago is representative of many other large cities, %as described in~\ref{chicago}, 
we lack the empirical evidence needed to ``assess the implications and potentially transformative consequences" of how similar smart city networks would emerge in different urban contexts~\cite{townsend2013smart}. An additional limitation is that we use weather data from US government agencies and there are only three weather stations in the Chicago area. Although we also had temperature and humidity readings at each node, these sensors were located inside the node enclosures, and thus did not always provide accurate external measurements. Thus, our weather-related analyses are not hyperlocalized to most of the sensors, and it is possible that there are hyperlocal weather correlations, such as urban heat islands, that affected sensor connectivity.

\section{Conclusion}
In this work, we present the challenges and opportunities from a year-term city-wide urban sensor network deployment. The network was created based on five specific criteria of success that we identified from past work. We provide an in-depth analysis of deployment data from the aspect of cellular connectivity and solar energy harvesting, which are the two key features that help meet the success criteria. In addition we highlight inherent challenges with open data sources available for root-cause analysis of failure nodes, and identify strengths and weaknesses to define future research directions that will support large-scale, real-time energy harvesting deployments in achieving reliable, equitable smart city networks.
% In this work, we present five criteria for successful urban sensor networks and analyze LTE connectivity and solar energy against these criteria through a year-long, citywide deployment. We highlight areas of opportunity for these technologies in helping to achieve reliable, equitable smart city networks. We determine that cell tower location, connectivity information, and land use data are essential for future urban sensor network planning, and encourage researchers to continue deploying networks in urban areas to identify other technical challenges in achieving smart urban futures.

\bibliographystyle{acm}
\bibliography{reference}

\end{document}